\documentclass[fleqn,10pt]{wlscirep}
\usepackage[utf8]{inputenc}
\usepackage[T1]{fontenc}
\usepackage[font=small,labelfont=bf,justification=justified]{caption}
\usepackage{multicol}

\usepackage{lastpage}
\usepackage{fancyhdr}
\usepackage{multibib}
\newcites{New}{References}

\title{Topological flocking models in spatially heterogeneous environments}

\author[1,2]{Parisa Rahmani}
\author[3,4]{Fernando Peruani}
\author[2,5,*]{Pawel Romanczuk}

\affil[1]{\scriptsize Department of Physics, Institute for Advanced Studies in Basic Sciences (IASBS), Zanjan 45137-66731, Iran}
\affil[2]{\scriptsize Institute for Theoretical Biology, Department of Biology, Humboldt Universit\"at zu Berlin, Berlin, Germany}
\affil[3]{\scriptsize Laboratoire de Physique Th{\'e}orique et Mod{\'e}lisation, CNRS UMR 8089, CY Cergy Paris Universit{\'e}, F-95302 Cergy-Pontoise Cedex, France}
\affil[4]{\scriptsize Laboratoire J.A. Dieudonn\'e, Universit\'e C\^ote d'Azur, CNRS UMR 7351, Parc Valrose, F-06108 Nice Cedex 02, France}
\affil[5]{\scriptsize Bernstein Center for Computational Neuroscience, Berlin, Germany}
\affil[*]{\scriptsize corresponding author: Pawel Romanczuk (pawel.romanczuk@hu-berlin.de)}

\begin{abstract}
Flocking models with metric and topological interactions are supposed to exhibit distinct features,  as for instance the presence and absence of moving polar bands.  On the other  hand,  quenched  disorder (spatial  heterogeneities) has been shown to dramatically affect large-scale  properties  of  active  systems  with  metric  interactions, while the impact of quenched disorder on active systems with metric-free interactions has remained,  until now,  unexplored.  Here, we show that topological flocking models recover several features of metric ones in homogeneous media, when placed in a heterogeneous environment.  In particular, we find that order is long-ranged even in the presence of spatial heterogeneities, and that the heterogeneous environment induces an effective density-order coupling facilitating emergence of traveling bands, which are observed in wide regions of parameter space.  We argue that such a coupling results from a fluctuation-induced rewiring of the topological interaction network, strongly enhanced by the presence of spatial heterogeneities.
\end{abstract}

\begin{document}

\flushbottom
\maketitle

\thispagestyle{empty}

\renewcommand{\figurename}{Fig.}


\noindent \section*{Introduction}
Flocking is a fascinating self-organized phenomenon observed in a large number of artificial and biological systems~\cite{vicsek2012collective,marchetti2013hydrodynamics}, including bacterial swarms~\cite{zhang2010collective, sokolov2012physical, peruani2012collective, gachelin2014collective}, fish schools~\cite{calovi2018disentangling}, and sheep herds~\cite{ginelli2015intermittent}, among many other examples.  
The large-scale properties of these active systems crucially depend on the type of interaction neighborhood of the moving agents.  
%
%
Two fundamentally different types of interaction neighborhood have been explored, the so-called metric~\cite{vicsek1995novel} and topological ones~\cite{ballerini2008interaction}.  

In metric models, the neighborhood of a particle is defined via the Euclidean distance between the focal and the neighboring particles, 
and  the number of neighbors of the focal particle scales with the local particle density.  
As a result of the competition between velocity alignment among neighbors and noise-induced decoherence, 
metric flocking models undergo spontaneous symmetry breaking~\cite{marchetti2013hydrodynamics}. 
In ideally homogeneous media, the order that emerges in these non-equilibrium systems in two dimensions is long-ranged (LRO)~\cite{toner1998flocks,toner2005hydrodynamics}, 
giant density fluctuations~\cite{toner2005hydrodynamics,ramaswamy2003active} are observed, and the phase transition is characterized by the presence of high-order, high-density bands that move across the system~\cite{gregoire2004onset,chate2008modeling,solon2015phase,solon2015pattern,bertin2006boltzmann,ihle2011kinetic,caussin2014emergent}.  
The presence of spatial heterogeneities or quenched disorder (e.g. obstacles, inhomogeneous substrates, etc), ubiquitous in all experimental and real-world active systems~\cite{bechinger2016active}, dramatically affects the large-scale properties of metric flocking models.  
In scalar active matter, it was found that obstacles can lead to jamming, frozen states, and moving chains~\cite{Reichhardt2014a,Reichhardt2014b,Yang2017,reichhardt2018}. 
For vectorial active matter in the presence of quenched disorder, it was shown first numerically~\cite{chepizhko2013optimal} and later analytically~\cite{toner2018swarming} 
that order becomes quasi-long-ranged (QLRO).  It was also found, using a minimal model,  there exists an optimal noise that maximizes collective motion~\cite{chepizhko2013optimal}, a result  later confirmed in more realistic simulations~\cite{martinez2018collective}. 
Furthermore, it was also predicted that spontaneous particle trapping leading to anomalous transport can occur~\cite{chepizhko2013diffusion}, a prediction in line with recent findings in bacteria~\cite{bhattacharjee2019bacterial}.  
In addition, it was also shown the existence of multiple attractor for flocks flying through the same realization of quenched disorder, meaning that the fate and history of the flock is strongly dependent on the initial condition~\cite{peruani2018cold}. 
Finally, it was found  in models~\cite{chepizhko2015active} and  experiments~\cite{morin2017distortion}, that above a given  density of spatial heterogeneities, polar bands vanish.  

While metric flocking models have been successful in reproducing several real active systems, it has been suggested that animals interact with a specific number of neighbors, regardless of local density, and thus independently of the relative Euclidean distance between the individuals~\cite{ballerini2008interaction}.  
The large-scale properties of topological flocking models are believed to be fundamentally different from the ones of the metric counterparts. 
In particular, the phase diagram of these systems, so far only studied in homogeneous media,  does not seem to possess a coexistence region characterized by the presence of polar, traveling bands~\cite{ginelli2010relevance,peshkov2012continuous,chou2012kinetic}; Fig. \ref{fig:coupling}a, d. 
The absence of traveling bands has been attributed to an apparent lack of a density-order coupling. 
On the other hand, the impact of quenched disorder on active matter with topological interactions has, so far, not been addressed. 
    Here, we address this open question in active matter theory by studying how quenched disorder affects the emergent properties of topological flocking models using $k$-nearest neighbors ($k$NN) and Voronoi tessellation\cite{rahmani2020flocking}.  
   We find that topological models differ fundamentally from their metric counterparts by exhibiting long-range order even in the presence of heterogeneities. 
Furthermore, we observe that in topological models, spatial heterogeneities counter-intuitively facilitate the emergence of traveling, polar bands (Fig. \ref{fig:coupling}b,e; and Supplementary Movie 1 and Supplementary Movie 2), while such elongated structures are believed not to be present in homogeneous media~\cite{ginelli2010relevance,martin2021fluctuation}. Finally, we argue that  band formation is related to the emergence of an effective coupling between local density and local order (Fig. \ref{fig:coupling}c, f.)  due to local rewiring of the interaction network, that is strongly enhanced by the presence of spatial heterogeneities.  

Our study provides a comprehensive characterization of the large-scale properties of topological flocking models in heterogeneous environments. 
\textcolor{black}{The results reported here, together with those by Martin \emph{et al}~\cite{martin2021fluctuation},  
strongly suggest that the established knowledge on topological flocking models needs to be fundamentally revised. 
Specifically, our analysis  extends our understanding of topological interactions in active matter systems by showing that topological flocking models in complex environments behaves as metric ones in homogeneous media.}

\section*{Results}

\subsection*{Model}
We consider active particles moving at constant speed $v_0$ in a two-dimensional, heterogeneous environment with periodic boundary conditions. 
The heterogeneous environment is modeled by a random distribution \textcolor{black}{of "obstacles" which we also will refer to as quenched  disorder or spatial heterogeneities.} 
Each active particle interacts with its topological neighborhood (TN), which define the particle's local environment. We use two definitions of TN: i) the first $k$-nearest neighbor ($k$NN) objects, and ii) all objects in the first shell by performing a Voronoi tessellation. 
Note that neighboring objects include other active particles, as well as obstacles.  
The behavior of particles is different for TN objects corresponding to active particles and obstacles: particles align their velocity to that of neighboring active particles and move away from obstacles. 
The equations of motion of $i$-th particle are given by:
%
\begin{equation}
\label{eq:motion}
\dot{\textbf{x}}_{i} = v_0 \textbf{V}(\theta_{i})
\end{equation}
\begin{equation}
\begin{split}
\dot{\theta_{i}} = &{} g_i(n_{o,i})\Bigl[\frac{\gamma}{n_{b,i}}\sum_{j\in \mathrm{TN}}\sin(\theta_{j}-\theta_{i})\Bigl] +\frac{\gamma}{n_{o,i}}\sum_{s\in \mathrm{TN}}\sin(\alpha_{s,i}-\theta_{i})+ \eta \xi_{i}(t)
\end{split}
\label{thetadot}
\end{equation}
where dots on the left-hand side denote temporal derivatives, $\textbf{x}_{i}$ is the position of the particle, and $\theta_i$ encodes the moving direction of the particle given by $\textbf{V}(\theta_i) = (\cos(\theta_i), \sin(\theta_i))$. 
The first term in Eq. (\ref{thetadot}) describes the alignment of the particle with TN active particles, while the second term describes repulsion from TN obstacles. 
The symbol $\mathrm{TN}$ denotes the set of topological neighbors of particle $i$, including $n_{b,i}$ active particles and $n_{o,i}$ obstacles. 
The position of TN obstacles is given by $\textbf{y}_{s}$, and $\alpha_{s,i}$ denotes the angle, in polar coordinates, of the vector $\textbf{x}_{i}-\textbf{y}_{s}$.    
Note that ``obstacles"  are in fact areas that the active particles avoid by turning away from their center ($\textbf{y}_{s}$), which can be viewed as a soft-core repulsive interaction.   
\textcolor{black}{Finally, $\gamma$ is a constant and  $\xi_{i}$ is a delta-correlated, dynamic  noise such that $\langle \xi_{i}(t) \rangle=0$ and $\langle \xi_{i}(t) \xi_{j}(t') \rangle=\delta_{ij} \delta(t-t')$; $\eta$ is a constant that denotes the strength of the dynamic noise.}  
We studied two options for $g_i(n_{o,i})$ that lead qualitatively to the same results:  
(a)  $g_i(n_{o,i})=1$ for all values of $n_{o,i}$, and (b) $g_i(n_{o,i}) = {1-\Theta[n_{o,i}]}$ with $\Theta[n_{o,i}] = 1$ if $n_{o,i}>0$. 
The latter option of $g_i(n_{o,i})$ ensures that in the presence of obstacles, the active particle gives priority to obstacles, moving away from them, ignoring  other active particles. 
Since results are easier to interpret with this rule, and are qualitatively the same as those obtained with $g_i(n_{o,i})=1$, 
we illustrate the system behavior using the obstacle priority rule; results for $g_i(n_{o,i})=1$ can be found in Supplementary Figure S1.    
In the following, we fix $\gamma=1$, $v_0=1$, $dt=0.1$, and particle density $\rho_b=N_b/L^2=1$, with $N_b$ the number of active particles in the simulation box of linear size $L$ (see Methods for further details). 

Note, that we have studied the dynamics of the above model recently also in the context of collective information processing~\cite{rahmani2020flocking}.

\subsection*{\textcolor{black}{Dynamic noise vs. quenched disorder}} 
\textcolor{black}{The system considered here contains two sources of fluctuation that promote  misalignment 
among the active particles: the dynamic noise and the quenched disorder (i.e. the obstacle field). For vanishing dynamic noise -- i.e. in the limit of ``cold" active matter -- the initial condition and specific distribution of obstacles determine the temporal evolution of the flock, implying that the system is not ergodic~\cite{peruani2018cold}. 
By including a non-vanishing dynamic noise, the systems remains strictly speaking non-ergodic, however time average quantities over long time-intervals can become independent of the initial condition.  Furthermore, we can expect that quenched disorder realizations sharing the same statistical properties -- e.g. same density of randomly distributed obstacles -- lead to similar time average quantities, as occurs for flocking models with metric interactions~\cite{chepizhko2013optimal}.} 

\textcolor{black}{To disentangle the level of fluctuation resulting from dynamic noise and quenched disorder, we compare the polar order parameter -- defined as $r = \langle r(t)\rangle = \left\langle|\frac{1}{N_b}\sum_{i} \exp(i\theta_{i})|\right\rangle$ -- computed over different realizations of statistically identical disorders; Fig.~\ref{fig:quench_vs_dyn_vor} (see Supplementary Figure S2 for the corresponding plots of $k$NN, $k=6$). 
Note that the standard error of the mean, $r$, over disorder realizations (red vertical lines), is either smaller than or of the same order of the variance of the polar order $r(t)$ over time for a single disorder realization (black curves). 
This strongly suggests that the large-scale properties of the system are highly similar among disorder realizations that share the same statistical properties. 
Finally, it is worth mentioning that for a given disorder realization in a finite system, though order can emerge in a large number of directions, not all of them exhibit the same probability. 
}

\subsection*{Optimal noise and long-range order} 

As shown in Fig.~\ref{fig:orderDisorder} a-c,  the polar order parameter  is a monotonically decreasing function of the noise strength $\eta$ in homogeneous environments with vanishing obstacle density $\rho_o=0$, whereby $\rho_{o}=N_o/L^2$ and $N_o$ is the number of obstacles in the system. 
One of the most remarkable features of metric flocking models in complex environments, i.e. for $\rho_o>0$, is the non-monotonic functional form of the curve $r$ vs. $\eta$ that puts in evidence the  presence of an optimal noise that maximizes collective motion~\cite{chepizhko2013optimal}.  
This optimal noise is absent in topological flocking models with $k$NN interaction: the curve $r$ vs. $\eta$ decreases monotonically with $\eta$  for all tested values of $\rho_{o}$, as occurs in homogeneous media, see Fig.~\ref{fig:orderDisorder}b (see Supplementary Figure S3 for the transition plots of other $k$ values). 
The situation for Voronoi neighbors is rather different. By increasing obstacle density $\rho_{o}$ from zero, a weak maximum appears in $r$ vs $\eta$, see Fig.~\ref{fig:orderDisorder}c, which tends to become weaker by further increasing $\rho_o$. 
One possible explanation for the lower order observed at low noise values is the formation of moving, high-density clusters that are only weakly interconnected among them, see Fig.~\ref{fig:orderDisorder}d-f and Supplementary Movie 3. 
\textcolor{black}{As observed for active particles with metric interactions in heterogeneous media~\cite{chepizhko2013optimal,chepizhko2015active}, 
we also find for Voronoi interactions, that a small, yet
finite values of dynamical noise facilitate exchange of directional information between clusters. At the optimal noise value the different clusters merge into a band-like structure and the global orientational order becomes maximal. A further increase of dynamical noise leads then to a monotonous decrease in order.}
In short, the existence of an optimal noise in topological flocking models seems to be model dependent.

A fundamental difference between topological and metric flocking models in complex environments is observed at the level of the emergent order. 
Metric flocking models in homogeneous media display LRO, while in heterogeneous media, order was shown, first numerically~\cite{chepizhko2013optimal} and later by an RG argument~\cite{toner2018swarming}, to becomes QLRO:  the polar order parameter $r$ decays algebraically with system size.  
On the other hand, topological models in homogeneous media and non-vanishing $\rho_b=N_b/L^2$, with $N_b$ is the number of active particles, also exhibit  LRO~\cite{ginelli2010relevance,peshkov2012continuous}.
By keeping $\rho_{b}$ and $\rho_{o}$ constant, while increasing $N_b$ and $N_o$, we provide solid numerical evidence indicating that the polar order parameter $r$ converges towards a constant value in the thermodynamic limit for both $k$NN and Voronoi neighbors at low and high obstacle densities. Specifically,  $\lim_{1/N_b \to 0} r \to r_{\infty}$, with $r_{\infty}$ a non-vanishing constant;  Fig.~\ref{fig:fssa}.

\textcolor{black}{
This result can also be obtained by 
studying $\langle \bar{\mathbf{V}}(\mathbf{r})\bar{\mathbf{V}}(\mathbf{0})\rangle$, 
where $\bar{\mathbf{V}}(\mathbf{r})$ refers to the local, average velocity of active particles in position $\mathbf{r}$, that as expected for LRO converges to a non-zero value for $|\mathbf{r}|\to \infty$, see Supplementary Figure S4.
} 
\textcolor{black}{
In addition, we have also confirmed the robustness of the observed LRO with respect to variation in the particle density $\rho_b$ by simulating systems with a larger and smaller density, $\rho_b=1.5$ and $\rho_b=0.5$ respectively (see Supplementary Figure S5).
}

In short, topological flocking models in heterogeneous media exhibit LRO, in contrast to the QLRO reported for the metric counterpart. 
\textcolor{black}{We note that as discussed further below, we observe formation of large scale bands for a wide range of parameters, in particular for the $k$NN model with $k=6$. Thus the corresponding LRO results are obtained in the presence of such emergent spatial structures.}

\subsection*{Traveling polar bands}  

In metric flocking models in homogeneous media, the emergence of polar bands has been explained as the result of a coupling between local polar order $r_{\ell}$ and local density $\rho_{\ell}$~\cite{solon2015pattern} 
\textcolor{black}{(see Methods for details regarding calculation of $r_\ell$ and $\rho_\ell$)}. 
On the other hand, topological flocking models have been introduced as active models that lead to large-scale, collective motion independently of the local density of the active particles~\cite{ballerini2008interaction}. 
In short, it has been assumed that in topological flocking models the above-mentioned order-density coupling is not present. Thus, traveling polar bands are not expected to emerge, as illustrated in Fig.~\ref{fig:coupling}a, d 
for Voronoi and $k$NN neighbors in homogeneous media.  
Fig.~\ref{fig:coupling}b, e and Figs.5a, 6a show that, counter-intuitively, by introducing inhomogeneities in the system, i.e. for $\rho_o>0$, traveling polar bands spontaneously emerge in topological flocking models across a wide range of parameters (see also Supplementary Movie 1 and 2). 
Moreover, for $\rho_o>0$ an effective order-density coupling, not observed for $\rho_o=0$, is present using both, Voronoi and $k$NN neighbors (Fig.~\ref{fig:coupling}c, f). 
To quantify the emergence of traveling, polar bands \textcolor{black}{we introduced a density modulation parameter $\beta$, defined via the amplitude of the largest Fourier mode with finite wave number $q>0$ of the Fourier-transformed coarse-grained density field (see Methods for details).}
%
Fig.~\ref{fig:Bandk6}b-e indicates $\beta$ at different noise values $\eta$ and obstacle densities $\rho_o$ for $k = 1$ and $k=6$. 
For $k=1$, bands are observed only near transition point (orange and red regions). 
By increasing $k$ -- e.g. to $k=6$ -- and $\rho_o$, 
bands are observed for all $\eta$ values such that $\eta < \eta_{c,H}$, where $\eta_{c,H}$ is the critical value in homogeneous media (Fig.~\ref{fig:Bandk6}). \textcolor{black}{We have confirmed that band structures emerge also for larger $k$ (e.g. $k=12$) over a wide range of parameters, in particular different noise intensities also away from the order-disorder transition (see Supplementary Figure S6).}

A core finding is that for fixed values of $\eta$ and $k$, bands becomes more pronounced as the obstacle density $\rho_o$ is increased; Fig.~\ref{fig:Bandk6} (and Fig.~\ref{fig:BandVornoi} for Voronoi interaction).  
This means that counter-intuitively the spatial heterogeneities promote band formation, while in metric models they hinder the formation of bands \cite{chepizhko2013optimal}.  
It is worth clarifying that this does not mean that spatial heterogeneities promote polar order, which decreases as $\rho_o$ increases. However, the presence of obstacles induce, as explained below, a coupling between local density and local (polar) order that leads to band formation.   
An important observation is that the speed of bands is independent of $\rho_o$ -- i.e. of quenched disorder -- and set by the amplitude $\eta$ of dynamic noise (see Supplementary Figure S7a-d). 
As the density $\rho_o$ is increased, the number of active particles traveling in the bands 
diminishes, the disorder gas density increases, and as result of this, the global polarization of the system decreases. 
One important lesson to draw is that it is not possible to reduce the impact of spatial heterogeneities to a re-normalized dynamic noise, since this would imply that the band speed depends on $\rho_o$,  which, we show, it does not.

%

\section*{Discussion}

\textcolor{black}{How can spatial heterogeneities promote band formation? 
In the following, we argue that 
(local) rewiring of the underlying dynamical network leads to an effective density-order coupling. 
Our argument is based on the following observations: 
i) local (orientational) order is strongly regulated by the level of (local) rewiring facilitating fast exchange of orientational information between different sets of particles, ii) obstacles induce local rewiring, and 
iii) rewiring is strongly density dependent, i.e. at high densities it will occur highly localized in space. This, together with the two previous points results in an emergent coupling between local (particle) density and local order, a necessary condition for band formation.}

\textcolor{black}{
The first assertion can be shown using a simple model.  
Assume a finite system of $N_s$ spins that when not connected to each other obey $\dot{\theta}_s = \eta \xi_s(t)$. 
At a rate $\nu$ pick a pair $i$-$j$ of spins and connect them for a finite time during which $\dot{\theta}_i = \gamma \sin(\theta_j - \theta_i) + \eta \xi_i(t)$ and $\dot{\theta}_j = \gamma \sin(\theta_i - \theta_j) + \eta \xi_j(t)$; for details see Methods. 
In this simple model,  order -- i.e. $|\sum_s \exp(i\theta_s)/N_s|$ -- increases with rewiring rate $\nu$ (Supplementary Figure S8a).  
This non-spatial model serves to prove that local rewiring can promote order.  }

\textcolor{black}{
The next step of the argument is to understand that 
the spatial motion of agents implies rewiring. 
This is evident for diffusing spins with metric interactions, where order is enhanced at larger densities or by using larger diffusion coefficients~\cite{grossmann2016superdiffusion}. Here, both effects result in faster exchange of interaction partners. 
In actual flocking models, however, the situation is more complex since particle velocity is coupled to $\theta_i$ and it is not possible to control the rewiring rate $\nu$ --  defined as the inverse of the average time an edge survives in the dynamical network -- 
without affecting the dynamics of $\theta_i$. However, simulations performed with topological flocking models in small systems -- Supplementary Figure S8b and Methods -- allow us to show that (local) polar order and the rewiring rate $\nu$ increase with the density of active particles $\rho_b$. 
Furthermore, 
in the vicinity of obstacles, active particles are forced to 
modify their trajectories, which affects the distance to neighboring particles, and leads, in consequence, to rewiring. 
Fig.~\ref{fig:MI}a, d confirms that, as expected, $\nu$ 
increases with the density of obstacles $\rho_o$. 
Here, we note that a finite obstacle density introduces naturally a characteristic maximal metric length scale of rewiring of the order of average obstacle distance $1/\sqrt \rho_o$. 
} 

\textcolor{black}{Finally, to quantify the coupling between local 
density $\rho_{\ell}$ and local order parameter $r_{\ell}$ 
we calculate the mutual information MI$(\rho_{\ell}, r_{\ell})$ as a measure of non-linear correlation between $\rho_{\ell}$ and $r_{\ell}$ , i.e. we quantify how much knowledge we gain on $r_{\ell}$ by knowing $\rho_{\ell}$ (see Fig.~\ref{fig:MI}b,c,e,f). 
It is important to note that $r_{\ell}$ is by definition bounded to values smaller or equal to $1$. For certain parameter choices, $r_{\ell}$ saturates to almost $1$ for all $\rho_{\ell}$ values. This occurs for instance at low dynamic noise and low obstacles densities. 
In these situations, it is evident that $r_{\ell}$ is independent of $\rho_{\ell}$, 
and thus MI$(\rho_{\ell}, r_{\ell})$ adopts low values. 
Note that for $r_{\ell} \sim 1$ particles are highly aligned and the relative distance between particles remains relatively constant implying a low rewiring rate. 
On the other hand, in the disordered state we observe $r_\ell\approx0$ for all $\rho_\ell$. Overall, in order for $r_{\ell}$ to be dependent on $\rho_{\ell}$, the system cannot be too disordered but also the level of polar order in the system should not be too high. 
This suggests, as actually observed, that sharper bands are observed at larger obstacle densities, where the level of global order is lower (see Supplementary Figure S7e). 
In particular, for a fixed dynamic noise, MI$(\rho_{\ell}, r_{\ell})$ 
is higher, indicating stronger correlation between $r_{\ell}$ and $\rho_{\ell}$,
for larger obstacle densities $\rho_o$, where the rewiring rate $\nu$ is also higher; see Fig.~\ref{fig:MI} and compare band snapshots in Figs.~\ref{fig:Bandk6} and \ref{fig:BandVornoi}. }

An interplay between $r_{\ell}$ and $\rho_{\ell}$ mediated by rewiring is arguably 
also present in spatially homogeneous systems. 
For fixed dynamic noise, it is expected that $\nu$ decreases with $k$ and increases with particle density $\rho$. Both trends are straight forward to understand under the assumption that the level of positional fluctuations of the particles are set by dynamic noise. For large values of $k$,  positional fluctuations only lead to replacement of the farthest neighbors, and this implies that most links (those corresponding to closer neighbors) are long lived. In consequence, the average time an edge survives increases with $k$, and its inverse, $\nu$, decreases. 
On the other hand, at high densities the average inter-individual distance between particles is small, and for the same level of positional fluctuations a higher rewiring rate is expected. 
Simulations performed in an homogeneous medium, Fig.~\ref{fig:homog}, are consistent with these arguments. 
In addition, all this suggests that the coupling between $r_{\ell}$ and $\rho_{\ell}$ in homogeneous media should be particularly strong for small $k$ values in the vicinity of the onset of order. 
Fig.~\ref{fig:homog}c shows that in an homogeneous medium for $k=1$ traveling bands robustly emerge, whereas they become quickly more diffuse with increasing $k$ and at $k=6$ are not observable in our simulations anymore  (see bottom panels in Fig. \ref{fig:Bandk6}a). This finding provides additional support for our arguments.
At this point, we also would like to point the reader to a recent publication~\cite{martin2021fluctuation}, which we learnt about at the time of submission, providing  
an alternative explanation for band formation in homogeneous media of flocking models with topological interaction. We note that the different mechanisms are not mutually exclusive in facilitating band formation in active matter with topological interactions.

\section*{Conclusions}

\textcolor{black}{We performed a comprehensive  study of  flocking models with topological interactions in heterogeneous environments. 
We investigated two different types of topological interactions, $k$NN and Voronoi, which are  
the two most studied active topological models in the literature~\cite{ballerini2008interaction,ginelli2010relevance,peshkov2012continuous,chou2012kinetic,martin2021fluctuation}. 
Similarly to what occurs in equilibrium system, where only few macroscopic details affect the emergent macroscopic behavior, here we found that the large-scale properties of these systems do not depend on the details of the implementation of the model -- e.g. on the choice of $g_i(n_{o,i})$  -- but on the nature of these interactions: i.e. topological (metric-free) interactions of polar symmetry. 
In that sense, our results are generic and we expect them to apply to other variations of topological interactions, as for example the spatially balanced $k$NN-model\cite{camperi2012spatially}. 
The two main results of our work on flocking models with topological interactions in heterogeneous environments are: 
1) We found that in sharp contrast to metric models -- where we observe quasi long-ranged order (QLRO) in heterogeneous environments  -- for topological models, according to our numerical study and up to the systems sizes investigated, the order is long-ranged (LRO) in the presence of spatial heterogeneities.   
2) We  showed that spatial heterogeneities promote the emergence of an effective density-order coupling that  allows active particles with topological interactions to form traveling polar bands, which share similar features to the bands observed in metric models. 
Importantly, bands are observed in parameter ranges in which metric models in homogeneous media do not develop bands. 
Furthermore, we argued that the counter-intuitive emergence of the density-order coupling for topological interactions is the result of the (local) rewiring of the underlying dynamical networks of active particles induced by the spatial heterogeneities. 
Finally, we expect that the numerical finding of LRO in heterogeneous media for  nonmetric active models -- arguably related to the presence of long-ranged connections between distant clusters -- 
will be supported by a RG calculation, as occurred in the past with the observation QLRO in metric models in the presence of quenched disorder~\cite{chepizhko2013optimal,toner2018swarming}.}

\textcolor{black}{In summary, our results show that topological flocking models in the presence of spatial heterogeneities -- which  
introduce a characteristic distance ($1/\sqrt{\rho_o}$)  --
behave as metric ones in homogeneous media, an observation that invites to a reconsideration of  ``metric-free'' interactions in active systems.}  

\section*{Methods}
\label{Methods}
\textcolor{black}{
\subsection*{Simulation details}
The model was implemented in C\texttt{++} programming language. Stochastic differential equations were solved using Euler–Maruyama method with an integration time step of $dt=0.1$. Topological interactions including $k$NN and Voronoi implemented using the CGAL computational geometry algorithm library \cite{cgal:eb-21a}. For $k$NN, $k$-d tree algorithm is used, where in order to account for periodic boundary conditions, the main simulation box has been repeated in different directions. In order to find particles in the first shell of Voronoi neighbors, the dual graph of Voronoi diagram i.e. (periodic implementation of) delaunay triangulation is used. 
}

\textcolor{black}{
All the other calculations and data processing have been done using Python and dependent libraries, in particular Numpy \cite{harris2020array} and Scipy \cite{scipy}.
}

\begin{figure}[]
\centering
\includegraphics[scale = 1.]{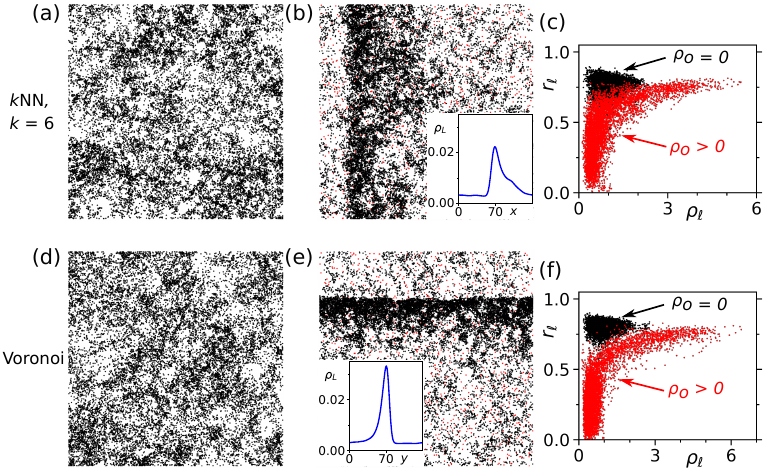}
\caption{\textbf{The role of heterogeneity in flocking with topological interactions}. Panels a, b, and c correspond to $k$-nearest neighbors interaction ($k$NN, $k=6$ where $k$ is the number of neighboring objects), and panels d, e, and f are for Voronoi interaction. Snapshots indicate macroscopic configurations, a, d) in obstacle-free environments, where obstacle density $\rho_o=0$, and b, e) in complex environments, with obstacle density $\rho_o=0.051$. Black and red dots represent particles and obstacles, respectively. The insets in b,e show 1D band profiles, that is 1D particle density $\rho_L$ along the moving direction. c, f) Local order ($r_\ell$) vs local density ($\rho_\ell$) measured in small cells of size $l=14$ in a simulation box with linear size 140 corresponding to the systems in previous panels (see Methods for details). Black scatters are for homogeneous, and red scatters are for heterogeneous environments. Noise intensity is fixed at $\eta=0.65$ here.
}
\label{fig:coupling}
\end{figure}

 \textcolor{black}{
\subsection*{Local density $\rho_{\ell}$ and local order $r_{\ell}$}
Density-order coupling plots of Figs.~\ref{fig:coupling} and \ref{fig:MI} have been obtained by superimposing the $r_{\ell}$ and $\rho_{\ell}$ of 60 snapshots taken from three time windows of simulations. In order to find these local quantities, the simulation box is divided into small cells of linear size $\ell$. Accordingly, local density is defined by $\frac{n_{\ell}}{\ell^2}$, where $n_\ell$ is the number of particles in the cell. And, local order is defined by $r_{\ell} = |\frac{1}{n_{\ell}}\sum_{i} \exp(i\theta_{i})|$, where the summation is over the $n_\ell$ particles of the cell. For the simulation box of size 140, we have used a cell size $\ell=14$. 
}

\textcolor{black}{
\subsection*{Quantification of bands}
{\it 1D band profile.--} In order to obtain 1D band profile, the density field of particles is smoothed using the kernel density estimation algorithm, then integrated over the direction perpendicular to the moving direction. Profiles represented in Fig. \ref{fig:coupling} are the result of averaging over 200 snapshots taken every 10 time steps.
}

\textcolor{black}{
{\it Band speed.--}
Speed of band is obtained by measuring the displacement of the peak of 1D band profile during a fixed time period.
}

\textcolor{black}{
{\it Band width.--}
Considering 1D band profile, band width is obtained from the difference between two points on the horizontal axis where the height of the profile is equal to a quarter of the maximum value.
}

\textcolor{black}{
{\it Density modulation parameter $\beta$.--} In order to quantify bands occurring in different regions of the parameter space, i.e. different $k$, $\rho_o$, and $\eta$, we cannot rely on band width obtained from 1D band profiles. Since, in addition to single bands, we also observe band-like density modulations or multiple bands, some of which are merging and splitting during the course of simulations. Therefore, obtaining a band-width which is representative of configurations of all the time steps, is in general not possible. In order to address this problem we use Fourier transformation of the coarse-grained density field and identify the maximal amplitude of the resulting Fourier spectrum for a finite spatial frequency $q>0$ (wave number). The density modulation parameter (maximum amplitude), $\beta$, is obtained after averaging over the power spectrum of 200 snapshots taken every 10 time steps. Please note that non-zero values of $\beta$ may also indicate other density modulation besides traveling bands, as for example formation of dense clusters in the Voronoi model for small dynamical noise (see Fig. \ref{fig:BandVornoi}). However, $\beta\gtrsim0.5$, typically indicate band formation.   
}

\textcolor{black}{
\subsection*{Order parameter fluctuations and error bars}
There are two kinds of fluctuations which affect the value of polar order parameter in our system. One stems from different realizations of obstacles in the environment, the so-called quenched disorder, the other is due to fluctuations in particles orientation, that is dynamic noise. The variation of the polar order $r(t)$ due to dynamic noise can be measured through its standard deviation, which will be correlated with the intensity of the dynamical noise.  
In the context of heterogeneous environments, the error of the (time-averaged) polar order parameter due to different realizations of the quenched disorder is the important quantity.
The error bars in Fig. \ref{fig:orderDisorder} and \ref{fig:fssa}, are calculated from 4 and 5 different realizations of random obstacle fields per parameter point, respectively. 
}

\textcolor{black}{\subsection*{From the non-spatial rewiring model to rewiring in small systems}
In order to show that rewiring can enhance (orientational) order,  we consider a simple non-spatial model. A system of small number, $N_s=10$, of spins with initial random orientations is considered. The system is such that at each time step there is only one link connecting two spins, $i$ and $j$. These two spins align with these rules, $\dot{\theta}_i = \gamma \sin(\theta_j - \theta_i)$ and $\dot{\theta}_j = \gamma \sin(\theta_i - \theta_j)$, while there is a random contribution $\eta \xi_s(t)$ to orientation of all the spins ($s$) in the system. The link between $i$, $j$ remains for $m$ time steps, then another two spins are selected randomly to interact. With this simple model, we show that smaller $m$, in other words, larger rate $\nu=1/t_m$ ($t_m=dt\cdot m$) of rewiring a single link results to a higher polar order $r$ for the system (see Supplementary Figure S8a). However, in flocking models, rewiring is associated to the relative motion of the particles, which in turn is related to the level of order. To verify that rewiring is correlated to the local level of order in the flocking model, we performed a series of small size simulations that clearly show such correlation between the rewiring rate $\nu$ -- which increases with the local density of particles as well as the density of obstacles -- and the level of order $r$; see Supplementary Figure S8b.}

\clearpage

\bibliographystyle{naturemag}
{\scriptsize
\bibliography{NatBib}}

\clearpage

{\scriptsize
\section*{\scriptsize Data availability}
The data-sets generated during the current study are available from the corresponding author on request.

\section*{\scriptsize Code availability}
The computer codes used for simulations and analyses are available from the corresponding authors upon request.

\section*{\scriptsize Acknowledgements}
P. Romanczuk acknowledges funding by the Deutsche Forschungsgemeinschaft (DFG, German Research Foundation) project number RO 4766/2-1. 
F. Peruani acknowledges support from the Agence Nationale de la Recherche via Grant No.~ANR-15-CE30-0002-01, project {\it BactPhys} and the Kavli Institute for Theoretical Physics (UCSB) and the organizers of the Active20 program for the online seminars and discussions. P. Rahmani acknowledges support from the Ministry of Science, Research and Technology of Iran (MSRT) and the German Academic Exchange Service (DAAD).

\section*{\scriptsize Author contributions statement}
P. Rahmani performed simulations. All the authors (P. Rahmani, F. Peruani, and P. Romanczuk) performed analysis, discussed results, and wrote the manuscript.


\section*{\scriptsize Competing interests}
The authors declare no competing interests.
}

\clearpage




\begin{figure}[]
\centering
\includegraphics[scale = 1.]{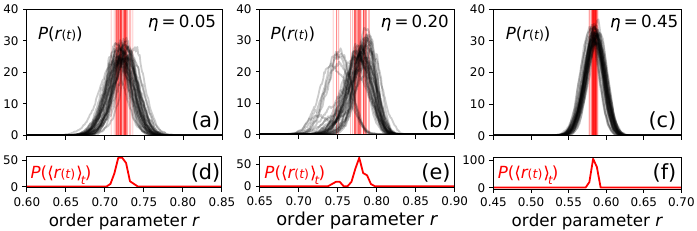}
\caption{\textbf{Dynamical noise vs quenched disorder}. Panels a, b, and c: Black curves correspond to probability density functions of polar order $r(t)$ obtained from stationary regime of 40 simulations with Voronoi interaction and different obstacle configurations, at obstacle density $\rho_{o}=0.15$ and noise intensity $\eta=0.05,0.20,0.45$ from left to right respectively. Red vertical lines correspond to mean (time-averaged) polarization of each of 40 distributions. Panels d, e, and f: Probability density function of the mean polarization values.
}
\label{fig:quench_vs_dyn_vor}
\end{figure}
 
\begin{figure}
\centering
\includegraphics[scale = 1.]{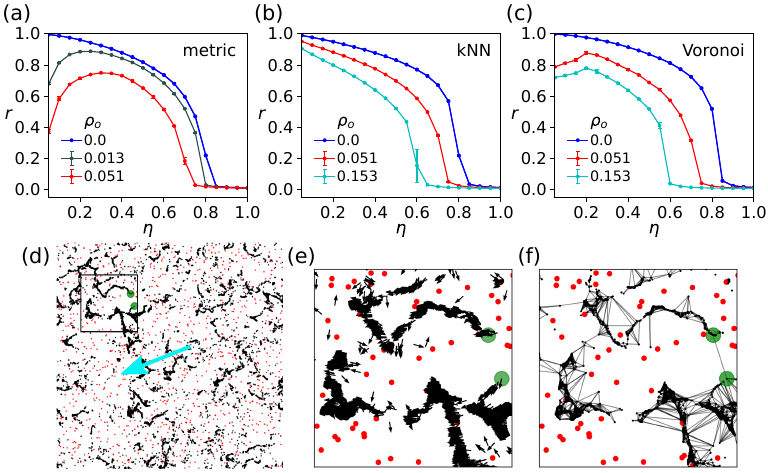}
\caption{\textbf{Disorder to order transition}. Polar order parameter $r$ versus noise intensity $\eta$ at different obstacle densities $\rho_o$ for different types of interaction neighborhoods, metric (a), $k$-nearest neighbors ($k$NN, $k=6$ where $k$ is the number of neighboring objects) (b), and Voronoi (c). d) The snapshot shows a typical configuration forming in a system with Voronoi interaction in heterogeneous environments at low noise values, here $\eta=0.05$, and $\rho_o=0.051$. Black and red dots represent particles and obstacles respectively. The arrow shows the instantaneous polarization of the system. 
(e) and (f) show magnification of the region displayed in (d). In (e), black arrows show the particles' instantaneous moving direction. In (f) the underlying (undirected) Voronoi interaction network is depicted. A link exists between two particles, if they are neighbors in the Voronoi tessellation, and if at least one of them does not have an obstacle in its neighborhood. The green circles point out clusters that are connected by
long links, i.e. long distance interactions. Error bars represent the standard deviation of the polar order parameter over different realizations of the obstacle field (see Methods). Note that error bars are often comparable or smaller than the symbol size.}
\label{fig:orderDisorder}
\end{figure}

\begin{figure}
\centering
\includegraphics[scale = 1.1]{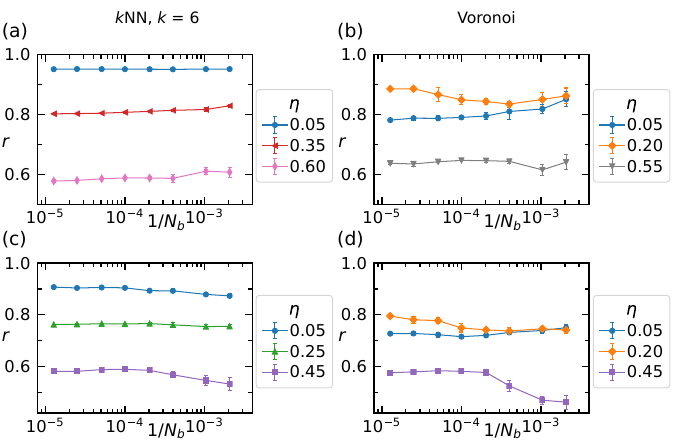}
\caption{\textbf{Probing the existence of long-range order}. Polar order parameter $r$ vs inverse system size $1/N_b$ at different noise values ($\eta$) and obstacle densities ($\rho_o$). Panels a, c correspond to $k$-nearest neighbors interaction ($k$NN, $k=6$ where $k$ is the number of neighboring objects), and panels b, d are for Voronoi interaction. a, b) $\rho_o = 0.051$ and c, d) $\rho_o = 0.153$. Error bars represent the standard deviation of the polar order parameter over different realizations of the obstacle field (see Methods). Note that error bars are often comparable or smaller than the symbol size.}
\label{fig:fssa}
\end{figure}

\begin{figure}[]
\centering
\includegraphics[scale = 0.6]{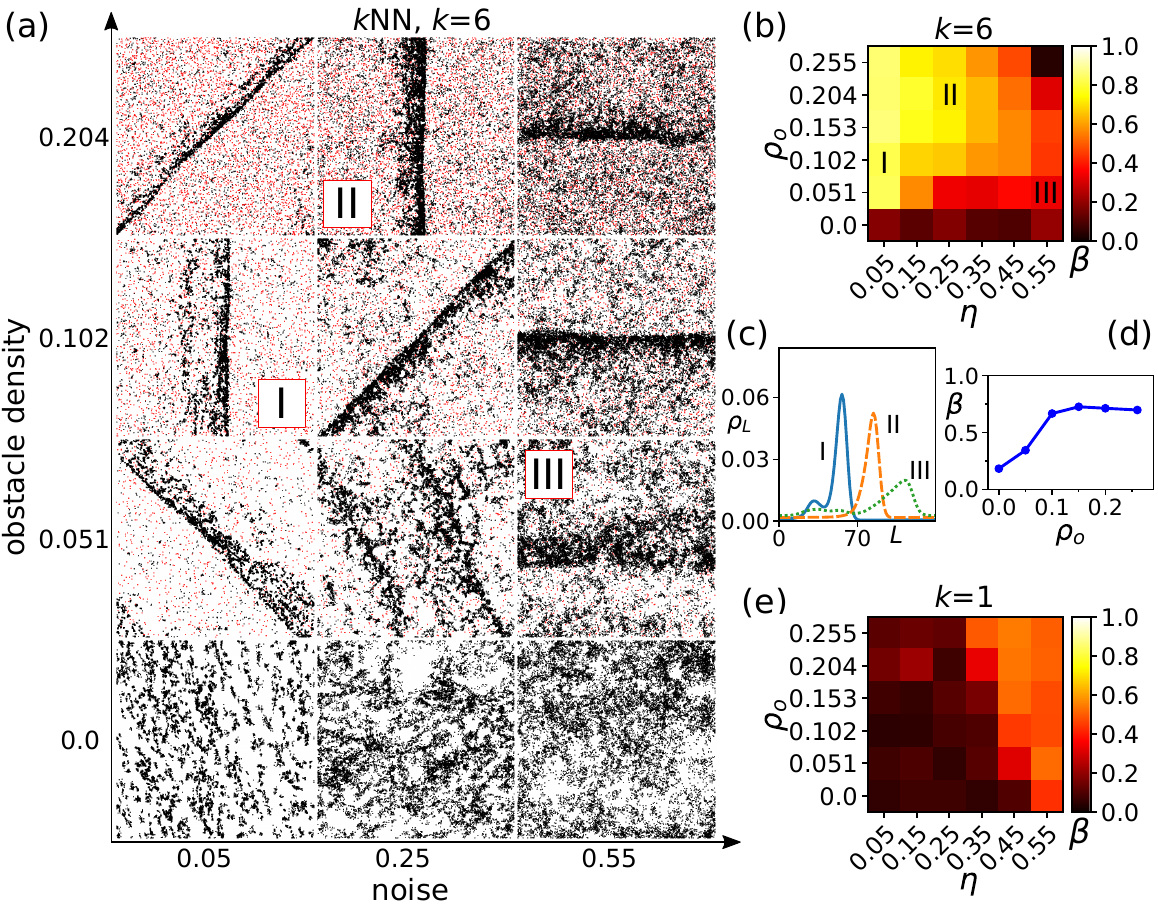}
    \caption{ \textbf{Emergent bands in $k$-nearest neighbors interaction ($k$NN)}. a) Typical macroscopic configurations observed in a system with $k$-nearest neighbor interaction ($k$NN, $k=6$, where $k$ is the number of neighboring objects), for different obstacle densities $\rho_o$, and noise intensities $\eta$. Black dots are particles and red dots are obstacles. In obstacle-free environments ($\rho_o=0$), we observe rather homogeneous structures with no bands. As we introduce obstacles, band-like structures emerge; at a fixed noise, by increasing obstacle density bands become sharper. For a fixed obstacle density, sharpest bands are observed at low noise values. Snapshots specified by I, II, and III have been used to calculate 1D band profiles in panel c.  b) Density modulation parameter $\beta$ in different regions of phase space specified with obstacle density $\rho_o$ and noise intensity $\eta$ corresponding to the system in panel a. 1D band profiles related to I, II, and III are shown in c. c) 1D band profiles - 1D particle density $\rho_L$ along the moving direction $L$ ($x$ or $y$) -  corresponding to the specified regions of panels a and b (regions I, II, and III). d) $\beta$ vs obstacle density $\rho_o$, showing the promoting effect of obstacles in band formation, corresponding to $\eta=0.25$ in panel b. e) Density modulation parameter $\beta$ for $k = 1$.
}
\label{fig:Bandk6}
\end{figure}

\begin{figure}[]
\centering
\includegraphics[scale = 0.7]{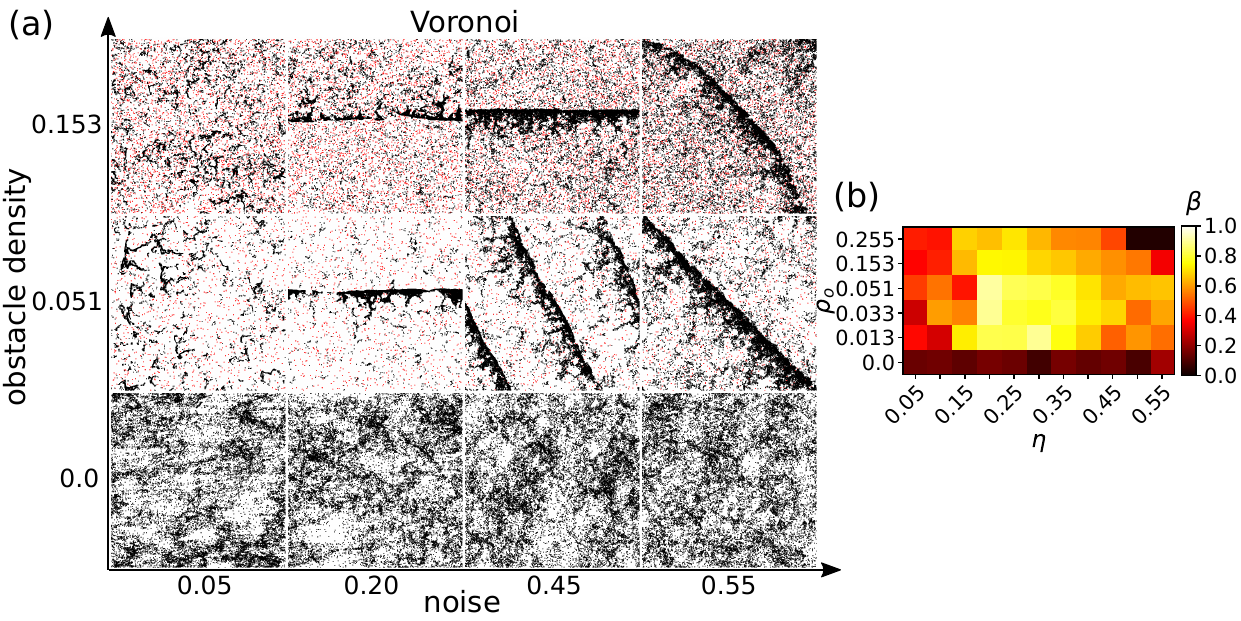}
\caption{\textbf{Bands in Voronoi interaction}. a) Typical macroscopic configurations observed in a system with Voronoi interaction, for different obstacle densities $\rho_o$, and noise intensities $\eta$. Black dots are particles and red dots are obstacles. In obstacle-free environments ($\rho_o=0$), we observe rather homogeneous structures with no density segregation in the form of cluster or band. In heterogeneous environments, we observe cluster-like structures at low noise values $\eta = 0.05$, these clusters connect together to form bands at noise around $\eta = 0.20$ and lead to a maximum in polar order parameter (see Fig. \ref{fig:orderDisorder}c). Bands continue to form at higher noise values. b) Density modulation parameter $\beta$ in different regions of phase space specified with obstacle density $\rho_o$ and noise intensity $\eta$ corresponding to the system in panel a. It should be noted that since macroscopic structures for Voronoi interaction are different from those with $k$-nearest neighbors interaction ($k$NN, where $k$ is the number of neighboring objects), with clusters at low noise values and bands at higher noise values, the same color (i.e. reddish regions) may refer to different types of structures when comparing to $k$NN interaction, and even for Voronoi when comparing high and low noise values.}
\label{fig:BandVornoi}
\end{figure}

\begin{figure}
\centering
\includegraphics[scale=0.95]{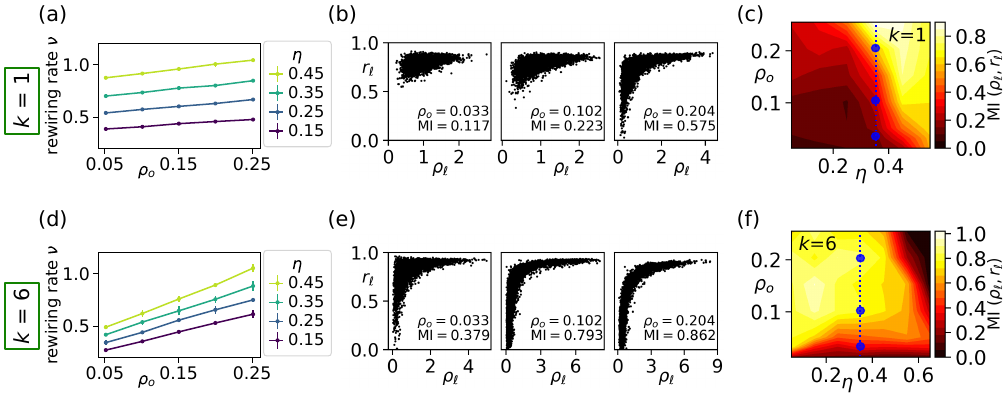}
    \caption{\textbf{Quantifying rewiring and the (local) density-order coupling in the $k$-nearest neighbor interaction ($k$NN).} The number of neighbors is $k$ = 1 (a, b, c)
and $k$ = 6 (d, e, f). a, d) The rewiring rate $\nu$ as a function of obstacle density $\rho_o$ for different values of dynamical noise. Error bars represent the standard deviation of rewiring over different realizations of the obstacle field. Note that error bars are often comparable or smaller than the symbol size. b, e) Local order $r_\ell$ versus local particle density $\rho_\ell$ for different obstacle densities $\rho_o$ (see Methods for details); the corresponding value of mutual information MI$(\rho_\ell,r_\ell)$ is also reported. c, f) Mutual information MI$(\rho_\ell,r_\ell)$ versus dynamical noise and obstacle density. The dashed line indicates the dynamical noise strength $\eta=0.35$ used in (b, e), with points indicating the corresponding obstacle densities.}
\label{fig:MI}
\end{figure}

\begin{figure}
\centering
\includegraphics[scale = 1.]{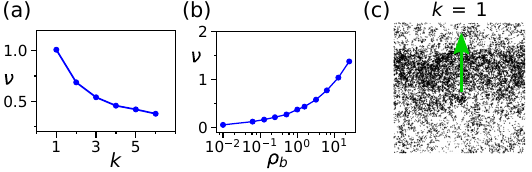}
\caption{\textbf{Observed bands for the $k$-nearest neighbor interaction ($k$NN) in homogeneous environments. $k$ is the number of neighbors and the density of obstacles is fixed at $\rho_o=0$}. a) Rewiring $\nu$ versus $k$ for a system with $N_b = 625$ particles (particle density $\rho_b=1.0$), interacting with $k$NN. b) Rewiring $\nu$ measured for the system with $N_b = 625$ at different particle densities $\rho_b$ by varying box size L ($k=6$). c) Snapshot showing band formation for $k$NN interaction with $k=1$. The black dots represent particles, and the green arrow shows moving direction of the band. Noise intensity is $\eta=0.55$ for all the panels.}
\label{fig:homog}
\end{figure}
\end{document}